\documentclass{pasj01}
\usepackage{url}
\usepackage{multirow}
\usepackage{graphicx}
\usepackage{lineno}
\usepackage{color}

\begin{document} 
\Received{2018/07/07}
\Accepted{2018/07/17}

\title{\ Late Engine Activity of GRB 161017A Revealed by Early Optical Observations}
\author{Yutaro \textsc{Tachibana}\altaffilmark{1}%
\thanks{}}
\altaffiltext{1}{Tokyo Institute of Technology, 2-12-1 Ookayama, Meguro, Tokyo 152-8551, Japan}
\email{tachibana@hp.phys.titech.ac.jp}

\author{Makoto \textsc{Arimoto}\altaffilmark{2}}
\altaffiltext{2}{Faculty of Mathematics and Physics, Institute of Science and Engineering, 
Kanazawa University, Kakuma, Kanazawa, Ishikawa 920-1192, Japan}

\author{Katsuaki \textsc{Asano}\altaffilmark{3}}
\altaffiltext{3}{Institute for Cosmic Ray Research, The University of Tokyo, 5-1-5 Kashiwanoha, Kashiwa, Chiba 277-8582, Japan}
\author{Shohei \textsc{Harita}\altaffilmark{1}}
\author{Taichi \textsc{Fujiwara}\altaffilmark{1}}
\author{Taketoshi \textsc{Yoshii}\altaffilmark{1}}
\author{Ryosuke \textsc{Itoh}\altaffilmark{1}}
\author{Katsuhiro L. \textsc{Murata}\altaffilmark{1}}
\author{Yoichi \textsc{Yatsu}\altaffilmark{1}}
\author{Kotaro \textsc{Morita}\altaffilmark{1}}
\author{Nobuyuki \textsc{Kawai}\altaffilmark{1}}

\maketitle
\begin{abstract}
The long gamma-ray burst GRB 161017A was detected by {\it Fermi} and {\it Swift},  
and its afterglow was observed by 
the {\it MITSuME} 50-cm optical telescope promptly about 50 s after the burst. 
Early optical observations revealed that the optical lightcurve exhibits a plateau and re-brightening
in the early afterglow phase about
500 and 5000 s after the trigger, respectively. 
By investigating the behavior of the spectral and temporal flux variation, 
it was found that the plateau and re-brightening cannot be explained 
in the context of the simple standard afterglow model. 
These observational features can be explained with two independent refreshed shocks, 
which indicate the long-acting central engine. 
We evaluated the physical parameters of the subsequent shells, 
and we then determined the kinetic energy ratio of the two colliding shells to the leading shell 
to be roughly 1 and 8, respectively. 
In addition, two prominent X-ray flares about 200 s after the trigger may be
signatures of delayed ejections of the energetic jets responsible for the refreshed shocks. 
Such late activity of the central engine and X-ray flares play a crucial role in understanding 
the mechanisms for jet formation and photon emission. 
\end{abstract}

\section{Introduction}
Gamma-ray bursts (GRBs) are the brightest events in the universe and exhibit
explosive electromagnetic emission from relativistic jets.
The emission episodes consist of a prompt gamma-ray emission and a temporally extended emission in the broad band from radio to gamma-ray band called ``afterglow''. 
 Pioneering X-ray observations of {\it HETE-2} \citep{2003AIPC..662....3R} enabled prompt localizations, alerts, and early optical afterglow observations by ground-based telescopes. 
The temporal behaviors observed from the lightcurves and the line features from the fine spectroscopy 
revealed the density profile of a circumstellar medium and the wind velocity around a progenitor to probe unknown characteristics of the massive progenitor and its environment, e.g., GRB 021004 \citep{2002A&A...396L...5L, 2003ApJ...588..387S} and GRB 030329 \citep{2003ApJ...597L.101T,2003ApJ...599L...9S, 2007ApJ...671..628T}. 
Furthermore, in the late phase of the optical afterglow, 
observations of several re-brightening features 
that drastically alter the temporal decay slope 
and do not restore the flux decay back to normal have been reported: 
e.g., GRB 021004 \citep{2003Natur.422..284F, 2004ApJ...615L..77B,2005A&A...443..841D} and GRB 030329 \citep{2003Natur.426..138G,2004PASJ...56S..77U}.
After the launch of the {\it Swift} satellite \citep{2004ApJ...611.1005G}, 
the instruments onboard {\it Swift} advanced the early afterglow in both the UV/optical and
X-ray bands and revealed complex temporal behavior of the lightcurves for many GRBs.

To explain the complexity, the model requires nontrivial assumptions such as a long-lasting internal shock activity, energy injection into the external forward-shock shell (e.g., \cite{2005Sci...309.1833B, 2006ApJ...639..316C, 2007PASJ...59..695A}), which is called the refreshed shock \citep{1998ApJ...496L...1R,1998ApJ...503..314P}, or a circumburst material with a wind profile \citep{2008ApJ...686.1209M, 2015ApJ...805...13L}. 
In some GRBs with more complex temporal behavior, a hybrid model such as a two-component jet model \citep{2007MNRAS.380..270O,2008Natur.455..183R,2011A&A...526A.113F} plus a reverse shock \citep{1997ApJ...476..232M,1999ApJ...520..641S,2000ApJ...545..807K} could be applicable. 

In particular, the continuous or discrete energy injection from the refreshed shock is related to the central engine activity
and could also probe photon emission caused by collisions between the external forward-shock shell and the late fresh shell.
The observational study of the refreshed shock is useful for understanding the dynamics of the system.

In this paper, we report multi-wavelength observations of GRB 161017A, focusing on the afterglow in the optical to X-ray band.
The optical light curves in the early afterglow phase exhibit a plateau and re-brightening about 500 and 5000 s after the trigger from {\it Swift}/BAT, respectively. We discuss their origin by examining two models: a reverse-forward shock and forward-refreshed shock. 
To explain the observed temporal and spectral evolutions in the optical band, the forward-refreshed shock is favored. If this interpretation is valid, the refreshed-shock feature at such an early phase with 
multi-color photometry will be a very rare and valuable sample to investigate the dynamics of GRB jets.

This paper is organized as follows. We present observations of GRB 161017A in Section \ref{Sec:Observation}. The obtained lightcurves and spectra are reported in Section \ref{Sec:Lightcurves} and Section \ref{Sec:sed}, respectively. We discuss the physical properties of the afterglow in Section \ref{sec:dis} and summarize our conclusions in Section \ref{sec:sum}.

\section{Observations}\label{Sec:Observation}
GRB 161017A triggered the {\it Swift}-BAT instrument at 17:51:51 UT on 2016 October 17 ($=t_0$),
 and {\it Swift}-XRT and UVOT detected the X-ray and optical counterparts at R.A., Dec. = 09$^\fh$31$^\fm$4$^\fs_.$53, +43$\fdg$07$\farcm$34$^\farcs_.$8 with an uncertainty of 1.4 arcsec (90\% containment; \cite{2016GCN..20064...1T}).
The {\it Fermi}-GBM instruments also detected the prompt emission of the burst simultaneously with the BAT \citep{2016GCN..20068...1H}. Its burst duration was 
$T_{\rm 90}$=32.3$\pm$8.1 s, where $T_{\rm 90}$ is the time interval over which 90\% of the total background-subtracted counts are observed. The obtained time-integrated spectrum in the prompt phase is well-represented by a power-law function with an exponential cutoff with a peak energy of $E_{\rm peak}$=299$\pm$45 keV, spectral index of $\beta = 0.0 \pm 0.1$, energy flux of (2.29 $\pm$ 0.14)$\times$10$^{-7}$ erg/cm$^2$/s, 
and fluence of $7.4 \times 10^{-6}$ erg/cm$^2$.
The redshift of the burst was determined to be z = 2.013 (\cite{2016GCN.20069....1D}). 

The MITSuME 50-cm telescope \citep{2005NCimC..28..755K, 2007PhyE...40..434Y, 2008AIPC.1000..543S, 2016RMxAC..48...24Y, 2017PASJ...69...63T} 
started multi-color optical observations 
(g$^{\prime}$, R$_{\mathrm C}$, and I$_{\mathrm C}$)
50 s after the trigger and detected a fading source \citep{2016Fuji}. 
Good sky conditions enabled MITSuME to obtain almost seamless data in the early afterglow phase until twilight 
($\sim$6$\times 10^3$ s after the trigger) 
with the exception of an interception from $\sim$4$\times10^3$ s to $\sim$5$\times 10^3$ s 
after the trigger by clouds. 
Further MITSuME multi-band observations were performed two days after the trigger. 
While {\it Swift}-UVOT also started optical observations $\sim$100 s after the trigger 
and detected a fading optical source (\cite{2016GCN.20074....1B}),  
UVOT cannot perform simultaneous multi-band observations; it needs to change filters. 
Thus, the MITSuME observation is useful for investigating the detailed temporal 
color variation in the early afterglow phase.

The raw data obtained by MITSuME were preprocessed in the standard manner -- subtracting dark and bias frames and then dividing by a flat frame. 
The detector pixel coordinates were calibrated into celestial coordinates using WCSTools (Mink 1997). After the primary treatment, 
we performed aperture photometry to estimate the magnitude of the GRB flux by comparing with three local reference stars using IRAF tasks after stacking some frames. 

\section{Lightcurves in Early Phase}\label{Sec:Lightcurves}
\subsection{Temporal Behavior}\label{sec:early}
\begin{figure}[t]
 \begin{center}
\includegraphics[width=8cm, bb = 11 4 479 342]{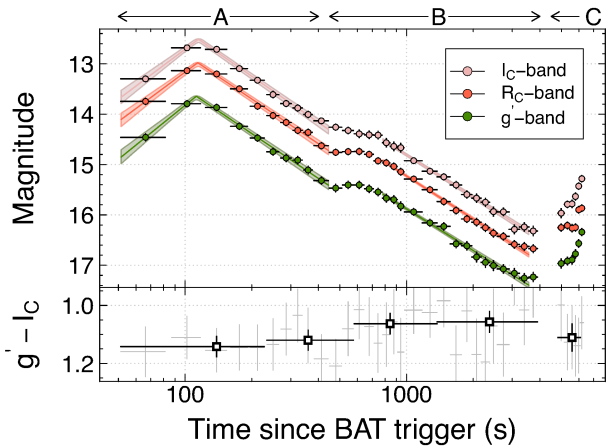} 
\end{center}
\caption{{\it Top}: Multi-color optical light curves of the early afterglow of GRB 161017A with three filters of g$^{\prime}$, R$_{\mathrm C}$, and I$_{\mathrm C}$. The shaded regions correspond to 1-$\sigma$ confidence regions from the temporal fittings shown in Table \ref{tab:par}. {\it Bottom}: (g$^{\prime}$ - $\mathrm{I_C}$) color curve.
The grey and black square points correspond to the single and binned data points of the above lightcurves, respectively. 
Smaller g$^{\prime}$ - $\mathrm{I_C}$ indicates bluer color (i.e., harder spectrum). }
\label{fig:lc}
\end{figure}
Figure \ref{fig:lc} presents the light curves of the early afterglow of GRB 161017A in three optical bands, which can be divided into three phases:
(A) a brightening (50 s $\leq t-t_0 \leq$ 100 s) and then decay until about 450 s; 
(B) a plateau (450 s $\leq t-t_0 \leq$ 650 s) and then decay until about 4 $\times10^3$ s; 
and (C) a rapid re-brightening (5$\times 10^3$ s $\leq t-t_0 \leq$  6$\times 10^3$ s). 
To analyze the temporal behavior quantitatively, 
we fitted a broken power-law function and a power-law function 
to the light curve of each filter (g$^{\prime}$, R$_{\mathrm C}$, and I$_{\mathrm C}$) in the time interval 
0 s $< t-t_0 <$450 s (entire phase A) and 750 s $< t-t_0 < $ 3500 s (only the decay part of  phase B)  
to obtain temporal indices.
In phase C, we tested both single and broken power-law functions to the lightcurves in the time interval of $t-t_0 > 10^4$ s. 

The parameters derived from the fits are listed in Table \ref{tab:par} for each filter. 
As all parameters are consistent in the three bands within the 1-$\sigma$ uncertainty in phase A and B, 
we performed a fit for the three optical bands 
with common temporal indices and peak times, 
while the normalization of each component is left unconstrained. 
The results are presented in the fourth and eighth rows of Table \ref{tab:par} 
(i.e., Filter: $g^{\prime}+R_{\mathrm{C}}+I_{\mathrm C}$). 
We hereafter use the values obtained by the common fit for the discussion of phase A and B.

\begin{table*}
\caption{Fitting results of the (broken) power-law function to the afterglow lightcurves of GRB 161017A. 
t$^{i}_{\mathrm{b}}$, $\alpha_{1}^{i}$, and $\alpha_{2}^{i}$ indicate the peak (or break) time and
the temporal indices before and after the peak (or break), respectively, where $i$ denotes the time intervals of A, B, and C. }
\begin{center}
\label{tab:par}
\begin{tabular}{llcccc}
\hline\hline
Time interval & Filter & $\alpha^{i}_{\mathrm{1}}$  & t$^{i}_{\mathrm{b}}$ & $\alpha^{i}_{\mathrm{2}}$ &  $\chi^2$/d.o.f.  \\
t$-$t$_0$ [s] &      &       & [s]    &     \\ \hline
(A) 50 -- 450  & g$^{\prime}$ & $+1.42\pm$0.16 & $(1.12\pm0.03)\times10^2$ & $-1.18\pm0.05$ & 4.1/6 \\ 
   & R$_\mathrm{C}$ & $+1.30\pm$0.15 & $(1.14\pm0.03)\times10^2$ & $-1.18\pm0.05$ & 4.8/6 \\ 
   & I$_\mathrm{C}$ & $+1.32\pm$0.15 & $(1.15\pm0.03)\times10^2$ & $-1.22\pm0.05$ & 6.5/6 \\  
   & g$^{\prime}$ $+$ R$_{\mathrm C}$ $+$ I$_{\mathrm C}$ & 
   $+1.38\pm$0.03 & $(1.13\pm0.02)\times10^2$ & $ -1.18\pm0.01$ & 19.2/24 \vspace{0.25cm} \\  \hline
(B) 750 -- 3500  & g$^{\prime}$  & $-1.09\pm0.03$ & - & - & 7.8/12    \\ 
\   & R$_{\mathrm C}$ & $-1.13\pm0.03$  & - & - & 5.8/12    \\ 
   & I$_{\mathrm C}$ & $-1.11\pm0.03$ & - & -  & 6.5/12    \\ 
   & g$^{\prime}$ $+$ R$_{\mathrm C}$ $+$ I$_{\mathrm C}$ & $-1.11\pm0.01$ & - & - & 32.2/38  \vspace{0.25cm} \\  \hline
(C) 10$^4$ -- 10$^6$  & R$_{\mathrm C}$+UVOT (model 1)  & $-1.38\pm0.02$ & - & - & 44.6/30    \\
   & R$_{\mathrm C}$+UVOT (model 2)  & $-1.23\pm0.07$ & $(3.4\pm1.3)\times10^4$ & $-1.47\pm0.07$ & 36.4/28    \\ 
     & X-ray$^{\ast}$ & $-1.24\pm0.09$ & $(4.6\pm1.0)\times10^4$ & $-1.82\pm0.11$   & 61.2/54  
\\ \hline
\end{tabular}
\end{center}
{\footnotesize ${}^{\ast}$ The parameters of the X-rays are obtained by fitting a broken power-law function
to the data after $t-t_0 = 10^4$ s to avoid the effect from the re-brightening. }
\end{table*}

\subsubsection{Phase A}
The brightening until $ t_0$ + 110 s probably corresponds to the onset of the afterglow.
The obtained temporal indices before and after the break ($\alpha_{1}^{\mathrm{A}}$ and $\alpha_{2}^{\mathrm{A}}$) 
are $+1.38\pm0.03$ and $-1.18\pm0.01$, respectively. 
The temporal rising index $\alpha_{1}^{\mathrm{A}} \sim +1.4$ is 
consistent with the distribution of the temporal rising index 
reported by other papers on the optical afterglows 
(e.g., $\alpha \sim$ 0.5--3; \cite{2010ApJ...723.1331M}). 
In contrast, the brightening in phase A is difficult to reconcile with 
the passage of the typical frequency of the forward shock emission; 
$\alpha \leq 0.5$ is predicted theoretically for the rising index (e.g., \cite{1999ApJ...520..641S, 2006ApJ...642..354Z}), 
and the optical color should be changed to redder by the passage 
(see Figure \ref{fig:lc} and Section \ref{sec:col} for color variation). 

\subsubsection{Phase B}
In phase B, a plateau from about $t_0 +$450 s to about $t_0 +$650 s is exhibited in the optical light curve. 
The obtained temporal decay index in the time interval 750 s $< t-t_0 <$ 3500 s ($\alpha_1^{\mathrm{B}}\sim-$1.11$\pm$0.01)
is slightly flatter than that in phase A ($\alpha_2^{\mathrm{A}} \sim -1.18\pm0.01$), 
and the optical color in this phase is bluer ({\it i.e.,} the spectral index is harder) 
than that in phase A, as represented in the color behavior of the bottom panel of Figure \ref{fig:lc}. 
The bluer trend after the plateau probably makes the flux increase in the I$_{\mathrm C}$ band less prominent 
than those in the R$_{\mathrm C}$ and g$^{\prime}$ bands.

\subsubsection{Phase C} \label{Sec:phC}
\begin{figure*}[t]
 \begin{center}
\includegraphics[width=13cm]{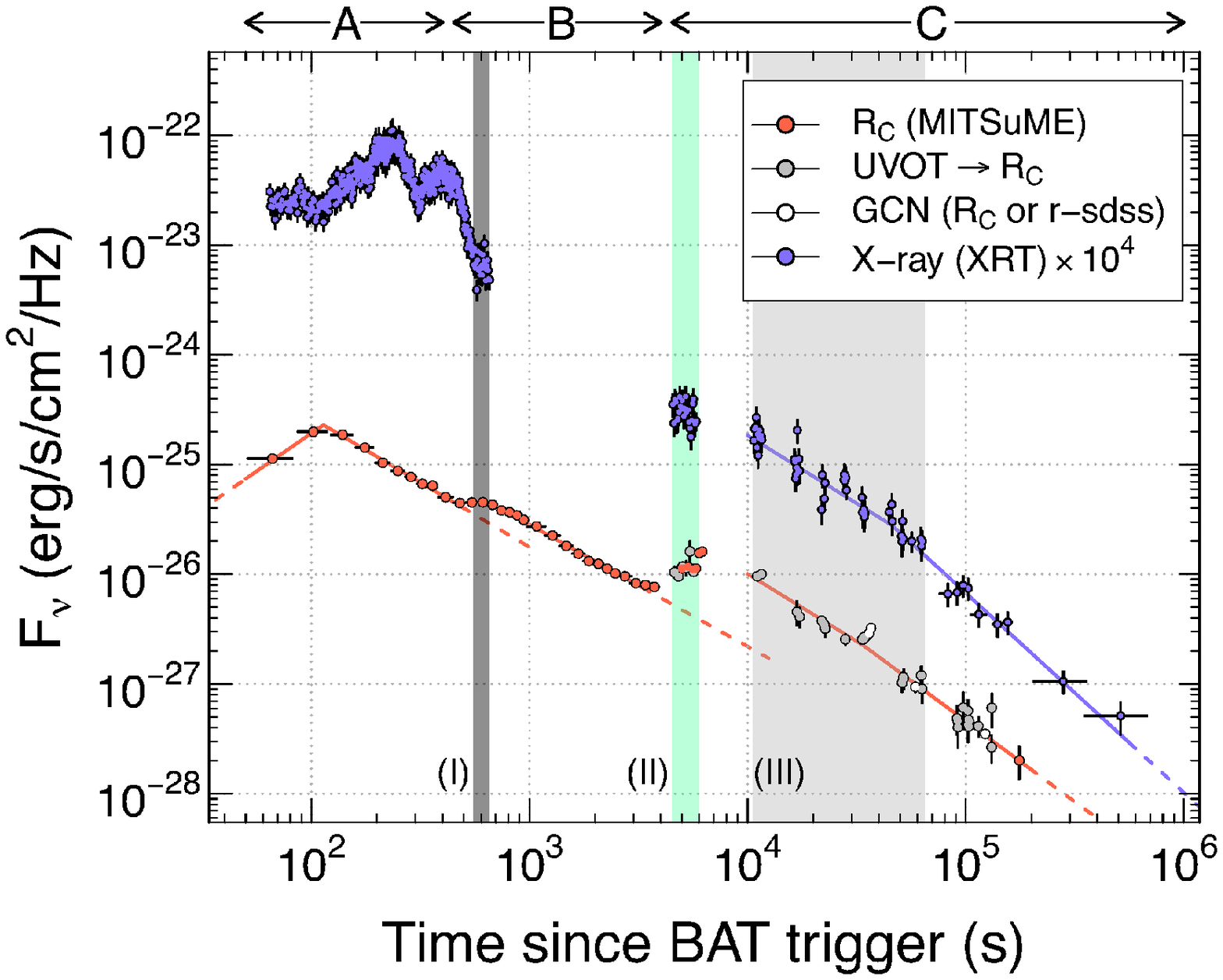}
\end{center}
\caption{Long-time behavior of the afterglow in GRB 161017A. 
The orange-filled circles represent the R$_{\rm C}$-band flux obtained by MITSuME, the
Grey-filled circles represent the photometric data from {\it Swift}/UVOT, 
and the white-filled circles represent the data reported from GCN 
(\cite{2016GCN.20071....1M}, \cite{2016GCN.20072....1G}, \cite{2016GCN.20080....1G}, and \cite{2016GCN.20079....1K}). 
The observed X-ray flux ({\it Swift}/XRT) is also represented by blue points for comparison. 
The shaded color regions denoted by dark grey, light green, and light grey correspond to the time intervals I, II, and III for the spectral analysis in Section \ref{Sec:sed}. 
The best-fit single and broken power-law functions for each phase (see Section \ref{Sec:Lightcurves}) are denoted by solid lines, 
and the extrapolated lines of each function are denoted by the dashed-line for reference. }
\label{fig:lc2}
\end{figure*}
At the beginning of phase C, a quite steep re-brightening occurs from $\sim$$t_0+$5000 s to $\sim$$t_0+$6000 s. 
From $\sim$6000 s after the trigger, MITSuME was no longer able to continue observation because of twilight. 
We therefore used {\it Swift}/UVOT data\footnote{\url{http://www.swift.ac.uk/burst_analyser/}} 
from $\sim$5000 s after the trigger 
to investigate how the optical afterglow evolves after this steep re-brightening. 
In Figure \ref{fig:lc2}, we present 
the long-term lightcurve with the {\it Swift}/UVOT flux densities 
converted to the 
R$_{\mathrm C}$-band measurements 
based on optical extinction and spectral index (see Section \ref{Sec:sed}).
We fitted a power-law function to the light curve from $10^4$ s 
and then obtained $\alpha_1^{\mathrm{C}} = -1.38 \pm 0.02$ with $\chi^2/\mathrm{d.o.f} \sim 45/30 = 1.5$. 
In contrast, when a broken power-law function was used for the fitting instead of a power-law function, 
we obtained $\alpha_1^{\mathrm{C}} = -1.23 \pm 0.07$, $\alpha_2^{\mathrm{C}} = -1.47 \pm 0.07$, 
and $t_{\mathrm {b}}^{\mathrm{C}} = (3.4 \pm 1.3) \times 10^4$ s 
with a better goodness of fit $\chi^2/\mathrm{d.o.f} \sim 36/28 \sim 1.3$ 
(the p-value of this improvement is about 5\%), 
where $t_{\mathrm {b}}^{\mathrm{C}}$ is the break time in the flux decay in phase C. 
The temporal decay index before the break of phase C ($\alpha_1^{\mathrm{C}}$) is 
consistent with that of phase A ($\alpha_2^{\mathrm{A}}$) within the 1-$\sigma$ confidence interval. 
In addition, the optical color likely returns to a similar value to that in phase A (Figure \ref{fig:lc}). 
Considering the improvement in the goodness of the fit, 
we adopt the parameters derived from the broken power-law function in the following discussion when referring to the optical behavior of phase C.

The X-ray lightcurve obtained by {\it Swift}/XRT is also presented in Figure \ref{fig:lc2} as a reference\footnote{\url{http://www.swift.ac.uk/xrt_live_cat/718023}}. In phase C, the X-ray temporal behavior is well-fitted by a broken power-law function, and the obtained parameters are listed in Table \ref{tab:par}.
Furthermore, we note that the break time ($t - t_0 \sim$ 5 $\times$ 10$^4$ s) and temporal decay index before the break ($\alpha_1^{\mathrm{C}} = -1.24 \pm 0.09$) 
in the X-ray band are consistent with those in the optical band within a 1-$\sigma$ uncertainty, although the temporal decay index after the break 
in the X-ray band ($\alpha_2^{\mathrm C} = -1.82\pm0.11$) is steeper than that in the optical band ($\alpha_2^{\mathrm C} = -1.47\pm0.07$).

\subsection{Color Variation}\label{sec:col}
In the bottom of Figure \ref{fig:lc}, we show the color variation 
(g$^{\prime} - $ I$_{\mathrm C}$) in the optical light curves. 
As already mentioned in Section \ref{sec:early}, 
a bluer trend is exhibited at the beginning of the plateau phase ($t - t_0$ $\sim$ 6 $\times10^2$ s) 
by $\Delta$ mag$_{g^{\prime} - I_{\mathrm C}} = -0.06\pm0.03$ between phase A and B 
\footnote{Changes in the color $\Delta$mag$_{g^{\prime} - I_{\mathrm C}}$ were 
calculated by the difference among three averaged colors in 
(A) 0 s $< t -t_0 <$ 450 s, (B) 750 s $< t -t_0 <$ 3500s, and (C) 4500 s $< t -t_0$.}. 
It then remained almost constant during phase B by comparing the color with the brightness. 
The color evolution corresponds to the change in the spectral index $\beta$ as 
\begin{equation}
\Delta \beta_{g^{\prime}, I_{\mathrm C}} 
= \frac{\Delta \mathrm{mag}_{g^{\prime} - I_{\mathrm C}}}{ \log_{\mathrm{10}} (\nu_{I_C}/\nu_{g^{\prime}})} 
\sim  -0.2\pm0.1, 
\end{equation}
where $\nu_{\mathrm{I_C}} = 3.7 \times 10^{14}$ Hz and $\nu_{\mathrm{g^{\prime}}} = 6.2 \times 10^{14}$ Hz \citep{1996Fuku}. 
After the beginning of the re-brightening ($\sim$6 $\times 10^3$ s; phase C), 
the color appears to have returned to almost the same level as that of phase A: 
$\Delta$ mag$_{g^{\prime} - I_{\mathrm C}} = -0.02\pm0.05$ between phase A and C, 
corresponding to $\Delta \beta_{g^{\prime}, I_{\mathrm C}}  = -0.1\pm0.2$. 
 
\section{SED from Optical-to-X-ray Band}\label{Sec:sed}
\begin{figure}[t]
 \begin{center}
\includegraphics[width=8.25cm, bb=4 24 478 483]{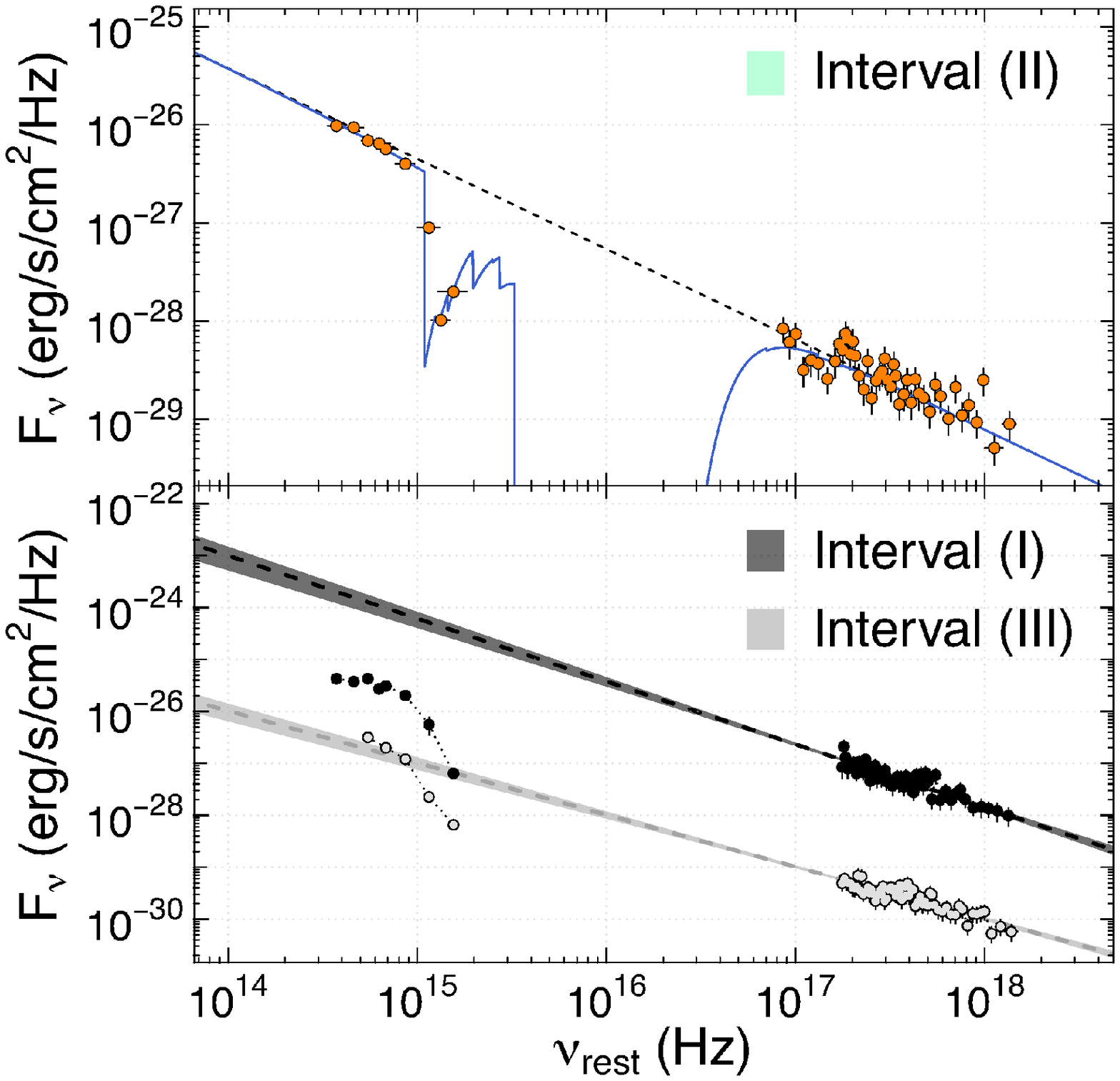} 
\end{center}
\caption{SEDs from the optical band to X-ray band for three time intervals:
(I) (6--7)$\times 10^2$ s, which is immediately after the flaring activity in the X-ray region (black points in the bottom panel), 
(II) (4--6)$\times 10^3$ s, which is where the re-brightening is taking place in the optical band (orange points in the top panel), 
and (III) (1--6)$\times10^4$ s, which is where the evolution of the X-ray light curve is similar to that in the optical band (grey points in the bottom panel), 
which are represented by the dark grey, light green, and light grey shaded color regions in Figure \ref{fig:lc2}, respectively. 
In the top panel, the blue solid line represents the best-fit model and the dashed line is the unabsorbed model. 
In the bottom panel, the dashed lines and shaded regions represent the best-fit power-law function 
and its 1-$\sigma$ uncertainty of the X-ray spectrum for each interval, respectively. }
\label{fig:sed}
\end{figure}

In Figure \ref{fig:sed}, 
we present spectral energy distributions (SEDs) for three time intervals: 
(I) $t-t_{0}\sim$ (5.5--6.5)$\times 10^2$ s, 
(II) $t-t_{0}\sim$ (4.5--6.0)$\times 10^3$ s, and 
(III) $t-t_{0}\sim$ (1.1--6.5)$\times 10^4$ s.

For interval (II), we performed a broadband SED fitting from the optical-to-X-ray band
by gathering g$^{\prime}$, R$_{\mathrm C}$, I$_{\mathrm C}$ (MITSuME), 
V, B, U, UVW1, UVW2, UVM2-band (UVOT), 
and the XRT data, with a model in {\tt XSPEC} \citep{1996ASPC..101...17A}; 
{\tt redden$\times$TBabs$\times$zdust$\times$zTBabs$\times$power}, 
where the redshift in {\tt zdust} and {\tt zTBabs} was fixed to 2.013.
The parameters derived by the fit are listed in the second row of Table \ref{tab:xpar}, 
where we assumed a Small Magellanic Cloud (SMC) environment with $R_{\mathrm V}=2.93$ for the host galaxy 
and $R_{\mathrm V} = 3.1$, $E(B-V)$ = 0.02 and 
$N_{\mathrm H} = 1.83\times10^{20}$ cm$^{-2}$ for our galaxy\footnote{\url{http://www.swift.ac.uk/xrt_spectra/00718023/}}. 
The fit is good for the optical and X-ray bands, 
enabling the optical and X-ray emission in this interval to be interpreted by a single emission region. 
In addition, we performed a model fit only to the optical-UV data. 
The derived parameters are listed in the third row of Table \ref{tab:xpar}.

When performing the spectral analysis for interval (I) and (III), 
we simply fit a power-law function only for the X-ray spectrum.  
The parameters derived by the fitting are also listed in Table \ref{tab:xpar}, 
where we fixed the $N_{\mathrm H}$ for the host galaxy 
to be $9.1 \times 10^{17}$ cm$^{-2}$, which was derived from the spectral fit for interval (II). 

\begin{table*}[]
\caption{Results of the spectral fit.}
\begin{center}
\scalebox{0.75}{
\begin{tabular}{lccccccccc}
\hline
Interval & redden & TBabs  & zdust  & zdust  &  zdust  & zTBabs & power-law & power-law & $\chi^2$/d.o.f.  \\
      & E(B$-$V) mag & N$_{\mathrm H}$ $\times 10^{20}$cm$^{-2}$ & method & E(B$-$V) mag & R$_{\mathrm V}$  & 
      N$_{\mathrm H}$  $\times 10^{17}$cm$^{-2}$& $\beta^{\ast}$ & Norm${}^{\ast \ast}$ &     \\ \hline
I (X-ray)    & -           & 1.83(fixed) & -      & -     & -           &
 9.1(fixed) & 1.21 $\pm$ 0.06& (1.2 $\pm$ 0.1) $\times 10^{-1}$ & 95/83\\
II (Opt. to X-ray)    & 0.02(fixed) & 1.83(fixed) & 3(SMC) & (5.5$^{+14.8}_{-5.5}$) $\times 10^{-3}$   & 2.93(fixed) &
 9.1 $\pm$ 0.3 & 0.92 $\pm$ 0.02 & (4.4 $\pm$ 0.2) $\times 10^{-3}$ & 66/50 \\
II (Opt. to UV)    & 0.02(fixed) & 1.83(fixed) & 3(SMC) & $<0.5$ & 2.93(fixed) &
 5.4 $\pm$ 0.6 & 1.15 $\pm$ 0.15 & ($9^{+14}_{-6}$) $\times 10^{-4}$ & 6/5 \\
III (X-ray)  & -           & 1.83(fixed) & -      & -     & -           & 
9.1(fixed) & 1.00 $\pm$ 0.05 & (6.4 $\pm$ 0.2) $\times 10^{-4}$  & 41/50 \\ \hline
\end{tabular}
}
\end{center}
{\footnotesize ${}^{\ast}$ Spectral index in the X-ray band.\\ 
${ }^{\ast \ast}$Power-law normalization at 1 keV (photons cm$^{-2}$ s$^{-1}$ keV$^{-1}$).}
\label{tab:xpar}
\end{table*}

In interval (I), the optical-to-UV flux is substantially below the power-law function extrapolated from the X-ray data, 
which would indicate that the X-ray emission is probably dominated by the emission from internal shocks
because the X-ray light curve exhibits a high variability. 
In this interpretation, the peak photon energy of the band function for this internal-shock component 
should be below the X-ray region at this time. 

In interval (III), 
the optical/UV SED is roughly within the 1-$\sigma$ error region of the power-law extrapolation from the X-ray data
(except the UVW1 and UVW2-band, where photons should be strongly absorbed by neutral hydrogen gas; 
the Lyman limit is located at about 1.1 $\times10^{15}$ Hz for z = 2.013). 
Although there must be an uncertainty in the optical-to-UV flux caused by the host galaxy absorption, 
the optical and X-ray emissions can be explained by the same origin in this period. 

\section{Discussion}\label{sec:dis} 
After the onset of the afterglow peak at $t-t_0\sim 110$ s, 
the MITSuME observation revealed 
the presence of a plateau phase starting at $t-t_0\sim500$ s, 
and a subsequent re-brightening is presented from $t-t_0\sim5000$ s. 
The temporal break at $t-t_0\sim $(3--5) $\times 10^4$ s occurs after the re-brightening. 
The temporal decay indices in each phase are obtained as 
$\alpha_{2}^{\mathrm{A}} = -1.18 \pm 0.01$, $\alpha_1^{\mathrm{B}} = - 1.11 \pm 0.01$, and 
$\alpha_1^{\mathrm{C}} = - 1.23 \pm 0.07$
by fitting the single or broken power-law function to the optical lightcurves in each phase, 
as described in Section \ref{Sec:Lightcurves}. 
We discuss the origin of this complicated behavior of the afterglow in the following section. 

\subsection{Single-component Jet?}\label{sec:single}

We first examine if the afterglow emissions in the optical and X-ray bands come from the same region, i.e.,
the single-component jet \citep{1997ApJ...476..232M}.

From the spectral point of view, the broadband SED can be well-fitted by a single power-law function 
from the optical-to-X-ray band in phase C, as shown in Section \ref{Sec:sed}. 
In addition, from the temporal point of view, 
the temporal decay index before the break at $t_{\mathrm b}^{\mathrm C}$ 
in the X-ray band is consistent with that in the optical band within a 1-$\sigma$ uncertainty:
$\alpha_1^{\mathrm{C}} = -1.23\pm0.07$ and $-1.24\pm0.09$ 
in the optical and the X-ray bands, respectively.
Furthermore, the temporal break time is consistent between the optical and X-ray bands, 
though the break time $t_{\mathrm b}^{\mathrm C}$ has a somewhat large uncertainty, as shown in Section \ref{Sec:phC}.
These results suggest that the emissions in the optical and X-ray bands have a common origin. 
Here, the simultaneous breaks in the optical and X-ray light curves at 
$t-t_0 \sim$ (3--5) $\times 10^4$ s are possibly a jet break; 
the temporal decay indices before/after the break in the X-ray band are consistent 
with those of the typical pre-/post-jet break, respectively (\cite{2009ApJ...698...43R}). 
Based on this interpretation, the opening angle of the jet $\theta_{\gamma}$ can be estimated by \citep{1999ApJ...519L..17S} 
\begin{equation}
\scalebox{0.8}{$\displaystyle
\theta_{\gamma} \simeq 
3.5\  \eta_{\gamma}^{1/8} 
	\left(  \frac{E_{\gamma, \mathrm{iso}}}{7.4\times10^{52}\ \mathrm{erg}} \right)^{-1/8}
	\left( \frac{n_{\rm ISM}}{1 {\rm \ cm^{-3}}}\right)^{1/8}
	\left( \frac{T_{\mathrm{b}}}{4\times10^4 \ {\rm s}} \right)^{3/8} \ \mathrm{deg}\ ,
$}
\end{equation}
assuming $z$ = 2.013, where $E_{\mathrm{\gamma, iso}}$ is the isotropic equivalent gamma-ray energy, 
$\eta_{\mathrm{\gamma}}$ is the radiative efficiency, 
$n_{\rm ISM}$ is the number density of the ISM, and
$T_{\mathrm{b}}$ is the jet break time of the afterglow in the observer frame 
(i.e., $T_{\mathrm b}=t_{\mathrm b}^{\mathrm C} \sim 4\times 10^4$ s). 
Assuming $\eta_{\mathrm{\gamma}} = 0.2$ and taking into account 
the increment in the kinetic energy of the leading shell by twice collisions 
(10 times larger than the initial $E_{\gamma, \mathrm{iso}}/\eta_{\gamma}$; see Section \ref{sec:kine}), 
the opening angle would be $\sim$2.1 deg. 
The obtained opening angle of the jet is in good agreement with
values derived from previous observations (e.g., \cite{2009ApJ...698...43R}). 

However, the temporal decay index after the break in the optical band ($\alpha_2^{\mathrm{C}} = -1.47\pm0.07$)
is slightly flatter than that in the X-ray band
($\alpha_2^{\mathrm{C}} = -1.82\pm0.11$). 
This inconsistency implies that another emission region,
({\it e.g.,} \cite{1998ApJ...499..301M}) may be present in the optical band, 
although there must be a systematic error in the temporal index 
because of the uncertainty in the flux conversion from the UVOT band to the $R_{\mathrm C}$ band. 
We therefore proceed to the discussion without the assumption that the afterglow emissions in the optical and X-ray bands are in the same region. 

\subsection{Reverse Shock in Phase A?}\label{sec:mec}
We investigate the physical origin of the afterglow emission in phase A. Immediately after the prompt emission, 
not only the forward shock (FS) emission but also the reverse shock (RS) emission 
would be expected in the optical band. 
As the theoretical prediction for the decay index of the RS expects 
$\alpha_{2}^{\mathrm{A}} = -(27p+7)/35$ for the constant-density interstellar medium (ISM), 
$\alpha_{2}^{\mathrm{A}} = -(p+4)/2$ for a stellar wind environment with a density profile $\propto r^{-2}$ 
(\cite{2000ApJ...545..807K, 2003ApJ...597..455K}), where $p$ is the electron spectral index. 
In the RS scenario, the electron spectral index $p$ in phase A must be quite hard as 
$p \sim +1.3$ and $-1.6$ for the ISM and the wind profile, respectively. 
The estimated indices appear to be too hard compared to the typical index $p \sim +2.1$ (e.g., \cite{1997ApJ...485L...5W, 2001ApJ...547..922F}),
 which is consistent with the theoretical estimate for the particle acceleration 
at relativistic shocks (e.g., \cite{2005PhRvL..94k1102K, 2013ApJ...776...46E}).
Therefore, the emission in phase A should not be dominated by the RS emission. 

\subsection{Forward Shock Scenario and the Circumburst Environment}\label{sec:mec}
As the optical observation does not favor the RS scenario, we consider the FS scenario. 
Furthermore, we estimate the electron spectral indices of the three FS components and the 
circumburst environment with the ISM or wind profile 
by examining both the temporal decay indices obtained from the lightcurve analysis and the spectral indices from the spectral analysis. 

Here, for the spectral analysis, we do not have the apparent spectral indices but do have the color variations of the afterglow in phase A and B. In contrast, we obtained a well-defined spectral index from the SED in phase C including intervals (II) and (III). As the color variations from phase A to C were already obtained, we can roughly estimate the spectral indices in phase A and B by an extrapolation from phase C, which is used in the following discussion. 

The obtained temporal decay indices of the optical lightcurves in phase A and C imply 
an electron spectral index of about
$2.6$ and about $1.9$ 
for the condition of $\nu_{\mathrm m} < \nu_{\mathrm{opt}} < \nu_{\mathrm{c}}$ 
in the ISM model and wind model, respectively, 
whereas $p \sim 2.2$ in the condition of $\nu_{\mathrm{c}} < \nu_{\mathrm{opt}}$ 
for both models (\cite{2006ApJ...642..354Z}). 
In contrast, the result from the spectral analysis in the time interval (II) expects 
$p \sim 2.5$ ($\nu_{\mathrm m} < \nu_{\mathrm{opt}} < \nu_{\mathrm{c}}$) 
or $p \sim 2.0$ ($\nu_{\mathrm{c}} < \nu_{\mathrm{opt}}$). 
Similarly, the optical color in phase A ($\Delta\beta = -0.1\pm0.2$ between phase A and C, as shown in Section \ref{sec:col}) 
leads to $p \sim$ 2.5 or $p \sim$ 2.0 
for $\nu_{\mathrm m} < \nu_{\mathrm{opt}} < \nu_{\mathrm{c}}$ or
$\nu_{\mathrm{c}} < \nu_{\mathrm{opt}}$, respectively. 
The value in the former case is consistent with $p \sim 2.6$ 
derived from the temporal decay index with the ISM model 
($\nu_{\mathrm m} < \nu_{\mathrm{opt}} < \nu_{\mathrm{c}}$) 
but is inconsistent with that in the wind model ($p \sim 1.9$). 
The fast-cooling regime ($p \sim 2.0$ for $\nu_{\mathrm c}<\nu_{\mathrm{opt}}$) is roughly consistent with the value 
derived from the optical temporal decay index: $p \sim 2.2$ 
with both the ISM and wind models. 

The spectral index in phase B is harder than those in the other phases 
($\Delta\beta = -0.2\pm0.1$ between phase A and B, as shown in Section \ref{sec:col}) 
and implies $p \sim$ 2.0 or $p \sim$ 1.5 
for $\nu_{\mathrm m} < \nu_{\mathrm{opt}} < \nu_{\mathrm{c}}$ or
$\nu_{\mathrm{c}} < \nu_{\mathrm{opt}}$, respectively. 
The former model appears more plausible, as the latter is characterized by a hard index, which could not be predicted easily using the standard shock theory. 
In addition, 
because a drastic change of $\nu_{\rm c}$ between phase A and B, or B and C is not likely, 
it should be reasonable to assume $\nu_{\rm m}<\nu_{\rm opt}<\nu_{\rm c}$ in the entire phase. 
Thus, considering the temporal and spectral features in the optical band in phases A, B, and C, 
the FS scenario can reproduce the observations, and, more specifically, the ISM model is favorable.

\subsection{Plateau Phase and Re-brightening} 
MITSuME clearly detected the plateau and re-brightening in the early afterglow phase, 
and the optical flux did not return to normal (see Figure \ref{fig:lc2}). 
For the origin of such behaviors in the afterglow light curves, 
the refreshed shock model is one of the most plausible scenarios (\cite{1998ApJ...496L...1R}). 
In this scenario, the variability timescale should be $\Delta t \ge t/4$ (\cite{2005ApJ...631..429I}). 
The variability timescales in the optical band are roughly $\Delta t_{\rm plateau} \sim 200$ s and $\Delta t_{\rm re} \sim 2000$ s 
for the plateau phase and re-brightening, 
and its starting times are $t_{\rm plateau} \sim 500$ s and $t_{\rm re} \sim5000$ s, respectively. 
The condition $\Delta t \ge t/4$ is thus satisfied in both cases. 

Flux increments for the plateau and re-brightening are about 1 and about 2.5 magnitudes from the flux 
extrapolated from the original baseline, respectively (Figure \ref{fig:lc}), 
which corresponds to a 2.5$\times$ and 10$\times$ brightening in each period. 
Such flux increments are similar to previously reported refreshed shocks in GRB afterglows 
({\it e.g.,} GRB 970508; \cite{1998ApJ...503..314P}, GRB 030329; \cite{2003Natur.426..138G}, 
GRB 070311; \cite{2007A&A...474..793G}, GRB 071003; \cite{2008ApJ...688..470P}, and
GRB 120326A; \cite{2014A&A...572A..55M}). 

The temporal decay index after the plateau $\alpha_1^{\mathrm{B}} = - 1.18 \pm 0.01$
is slightly shallower than that before the plateau $\alpha_{2}^{\mathrm{A}} = - 1.11 \pm 0.01$. 
As we mentioned in Section \ref{sec:mec}, the change in the decay index $\Delta \alpha \sim 0.1$ between phase A and B 
corresponding to the change in the electron spectral index $\Delta p \sim -0.15$ ($\Delta\beta \sim -0.1$)
is roughly consistent with that evaluated from the change in the optical color (Section \ref{sec:col}). 
In previous works, there were a few examples that showed the re-brightenings 
accompanied with different spectral changes in the optical band: 
harder when brighter for GRB 071031 (\cite{2009ApJ...697..758K}) 
and softer when brighter for GRB 081029 (\cite{2011A&A...531A..39N}). 
GRB161017A has a harder-when-brighter feature in phase B and a softer-when-brighter feature in phase C.
Although these emission mechanisms are still unclear, these different obtained results 
imply a diversity of electron acceleration properties and late engine activity such as refreshed shocks.

\subsection{Kinetic Energy and Emission Efficiency of the Shells}\label{sec:kine}
The multiplying factor in the optical flux $F$ against the extrapolation of the first component, 
$F_{\mathrm{after}}/F_{\mathrm{before}}$, is estimated to be about 2.5 and about 10
in the plateau phase and re-brightening phase, respectively. 
For the case of $\nu_{\mathrm m} < \nu_{\mathrm{opt}} < \nu_{\mathrm{c}}$, 
the ratio of the bulk kinetic energy after the shell merges to that of the leading shell 
can be roughly estimated as 
$(F_{\mathrm{after}}/F_{\mathrm{before}})^{4/(p+3)} \sim $ 2 and 5 (\cite{{2003Natur.426..138G}}), 
where we adopt $p=2$ and $p=2.5$ for the plateau and re-brightening cases, 
indicating that the kinetic energy ratio of the subsequent shells to the leading shell is about $1$ and 8, respectively. 

Figure \ref{fig:lc2} shows that while the prompt emission lasts until $t=40$ s, 
two prominent X-ray flares at $t \sim $ 200 and 400 s occur.
A simple interpretation could be that the optical emission in phase A is
mainly attributed to the ejecta responsible for the prompt emission, 
and the two subsequent shells then contribute to the plateau and re-brightening phases 
after they have emitted the X-ray flares at $t \sim $ 200 and 400 s, respectively.

The fluences for the two X-ray flares are estimated as 
$(1.36\pm0.01) \times10^{-6}$ erg/cm$^2$ and 
$(0.80\pm0.01) \times10^{-6}$ erg/cm$^2$, respectively. 
Those are 5--10 times lower than that of the prompt emission 
($7.4 \times10^{-6}$ erg/cm$^2$), 
though the kinetic energies of the two shells are 
comparable to
or larger than the energy of the primary ejecta, 
as we have estimated. 
If the above interpretation is valid, the photon emission efficiencies for the two X-ray flares 
should be much lower than the prompt emission efficiency.
In this case, even though the central engine ejected larger energies in the later phase, 
the photon emission efficiency decreases with time. 
This may imply a change in the photon emission mechanism, for instance, 
from the photosphere emission to the internal shock emission.

\subsection{Lorentz Factor of the Fast and Slow Shells}
In this section, we evaluate the physical parameters for each shell: 
the fast-leading shell that is interacting with the external material (ISM), 
and two slow shells that catch up with the fast-leading shell. 

The bulk Lorentz factor of the leading shell evolves as $\Gamma \propto T_{\rm obs}^{-3/8}$ \citep{1998ApJ...497L..17S}. 
Adopting $T_{\rm obs}=t_{\rm b}^{\rm A}$, we obtain $\Gamma$ at the onset of the shell deceleration as
\begin{equation}
\scalebox{0.8}{$\displaystyle
\Gamma_0 \simeq 145 \ \eta_{\mathrm{\gamma}}^{-1/8} \left( \frac{E_{\mathrm{\gamma, iso}}}{7.4 \times 10^{52} \ {\rm erg}}\right)^{1/8} \left( \frac{n_{\rm ISM}}{1 {\rm \ cm^{-3}}}\right)^{-1/8} \left( \frac{t_{\rm b}^{\rm A}}{110 \ {\rm s}} \right)^{-3/8}  \  {,}
\label{eq:ini}
$}
\end{equation}
assuming $z$ = 2.013, which can be regarded as the initial Lorentz factor\footnote{
The conventional formula for $\Gamma_0$, 
Eq.(10) in \cite{1999ApJ...520..641S}, may overestimate the value \citep{2017ApJ...844...92F}. 
We thus adopt Eq.(\ref{eq:ini}), which is consistent with the later estimates of $\Gamma$ for later period.
}. 
Hereafter we assume $\eta_{\gamma} = 0.2$, which leads to $\Gamma_0 \sim 180$. 

For the first slow shell associated with the plateau phase ($T_{\mathrm{obs}} \sim 500$ s), 
the shell should be slower than $\Gamma_0$ but faster than 
the Lorentz factor of the leading shell at $T_{\mathrm{obs}} \sim 500$ s. 
We therefore obtain the range of the Lorentz factor of the first slow shell $\Gamma_{\mathrm{s1}}$ as $100 < \Gamma_{\mathrm{s1}} < 180$. 
On the other hand, for the second slow shell attributed to the re-brightening phase ($T_{\mathrm{obs}} \sim 5000$ s), 
its Lorentz factor ($\Gamma_{\mathrm{s2}}$) can be estimated by $2\Gamma(T_{\mathrm{obs}})$ \citep{2000ApJ...532..286K} 
with the approximation of $T_{\rm obs} \gg t_{\rm b}^{\rm A}$. 
Taking into account that $E_{\gamma, \mathrm{iso}}/\eta_{\gamma}$ of the leading shell  
would be about two times larger after the collision with the first slow shell at the plateau phase (see Section \ref{sec:kine}), 
we can obtain the Lorentz factor of the second slow shell as $\Gamma_{\mathrm{s2}} \sim 90$. 

\section{Summary} \label{sec:sum}
The MITSuME telescope observed the early afterglow in GRB 161017A 
and clearly detected the onset of the afterglow, a plateau, and a re-brightening
at about 110, about 500, and about 5000 s after the trigger, respectively. 
We concluded that the afterglow is dominated by the FS propagating in an ISM environment 
with $\nu_{\mathrm m} < \nu_{\mathrm{opt}} < \nu_{\mathrm{c}}$, 
and the plateau and re-brightening can be interpreted by two independent refreshed shocks. 

The kinetic energy ratio of the subsequent shells to the leading shell are evaluated to be about 
1 and 8, respectively. 
The leading ejecta, which may be responsible for the prompt gamma-ray emission, 
 dominantly contributes to the early phase of the afterglow before the plateau phase. 
 In addition, the injected energies at the plateau and re-brightening phase are 
comparable to 
or larger than the kinetic energy of the initial shell. 
This implies that the central engine released more energy as jets with a lower photon-emission efficiency 
in the X-ray flare activity. 
This qualitative change in the activity 
may provide us with a hint to understand the mechanisms of jet formation and prompt/X-ray flare emission. 

Unfortunately, the lack of XRT coverage approximately between 700 s and 4 ks
and of the multi-color optical observation after about 10 ks from
the trigger makes it difficult to investigate the evolution of the
broadband spectrum.
Therefore, we could not strongly constrain the physical
origin of the plateau and the re-brightening in this study.
Further continuous efforts on multi-wavelength, more immediate, and
seamless observations of GRBs would boost our understanding of the GRB
emission mechanism. 


\begin{ack}
This work was supported by JSPS KAKENHI Grant Numbers JP16J05742 (Y.T.) and JP17H06362 (M.A. and N.K.). 
Y.T. is also financially supported by Academy for
Global Leadership (AGL) of Tokyo Institute of Technology. 
M.A. acknowledges the support from the JSPS Leading Initiative for Excellent Young Researchers program. 
This work was also supported by the joint research program of 
the Institute for Cosmic Ray Research (ICRR), the University
of Tokyo, Grants-in-Aid for Scientific Research nos.16K05291 (K.A.), 
JSPS and NSF
under the JSPS-NSF Partnerships for
International Research and Education (PIRE; R.I.), 
and 
Optical and Near-Infrared Astronomy Inter-University Cooperation Program (K.L.M.).

\end{ack}


\begin{thebibliography}{}
\bibliographystyle{apj}
\bibliography{GRB161017A}
\bibitem[Arnaud(1996)]{1996ASPC..101...17A} Arnaud, K.~A.\ 1996, 
Astronomical Data Analysis Software and Systems V, 101, 17 
\bibitem[Arimoto et al.(2007)]{2007PASJ...59..695A} Arimoto, M., Kawai, N., Suzuki, M., et al.\ 2007, \pasj, 59, 695 
\bibitem[Bj{\"o}rnsson et al.(2004)]{2004ApJ...615L..77B} Bj{\"o}rnsson, G., Gudmundsson, E.~H., \& J{\'o}hannesson, G.\ 2004, \apjl, 615, L77 
\bibitem[Breeveld \& Troja (2016)]{2016GCN.20074....1B} {Breeveld}, A.~A. \& {Troja}, E. \ 2016, GCNC, 20074
\bibitem[Burrows et al.(2005)]{2005Sci...309.1833B} Burrows, D.~N., Romano, P., Falcone, A., et al.\ 2005, Science, 309, 1833 
\bibitem[Cusumano et al.(2006)]{2006ApJ...639..316C} {Cusumano}, G., {Mangano}, V., {Angelini}, L., et al. \ 2006, \apj, 639, 316
\bibitem[Dai \& Lu(1998)]{1998A&A...333L..87D} Dai, Z.~G., \& Lu, T.\ 1998, \aap, 333, L87 
\bibitem[D'Avanzo et al.(2016)]{2016Avan} D'Avanzo, P., Malesani, D., D$^{\prime}$Elia, V., et al. \ 2016, GCNC 
\bibitem[de Ugarte Postigo et al.(2005)]{2005A&A...443..841D} de Ugarte Postigo, A., Castro-Tirado, A.~J., Gorosabel, J., et al.\ 2005, \aap, 443, 841
\bibitem[de Ugarte Postigo et al.(2016)]{2016GCN.20069....1D} {de Ugarte Postigo}, A. and {Kann}, D.~A. and {Thoene}, C., et al. \ 2016, GCNC 
\bibitem[Ellison et al.(2013)]{2013ApJ...776...46E} {Ellison}, D.~C. and {Warren}, D.~C. and {Bykov}, A.~M. \ 2013, \apj, 776, 46 
\bibitem[Filgas et al.(2011)]{2011A&A...526A.113F} Filgas, R., Kr{\"u}hler, T., Greiner, J., et al.\ 2011, \aap, 526, A113 
\bibitem[Fukushima et al.(2017)]{2017ApJ...844...92F} {Fukushima}, T.,  {To}, S. \& {Asano}, K., et al. \ 2017, \apj, 844, 92
\bibitem[Fox et al.(2003)]{2003Natur.422..284F} Fox, D.~W., Yost, S., Kulkarni, S.~R., et al.\ 2003, \nat, 422, 284 
\bibitem[Fukugita et al.(1996)]{1996Fuku} Fukugita, M., Ichikawa, T., Gunn, J. E., et al. \ 1996, \apj, 1748, 111
\bibitem[Gehrels et al.(2004)]{2004ApJ...611.1005G} Gehrels, N., Chincarini, G., Giommi, P., et al.\ 2004, \apj, 611, 1005 
\bibitem[Granot et al.(2003)]{2003Natur.426..138G} {Granot}, J., {Nakar}, E. \& {Piran}, T.\ 2003, \nat, 426, 138 
\bibitem[Guidorzi et al.(2007)]{2007A&A...474..793G} {Guidorzi}, C., {Vergani}, S.~D., {Sazonov}, S. \ 2016, \aap, 474, 793
\bibitem[Guidorzi et al.(2016A)]{2016GCN.20072....1G} {Guidorzi}, C., {Kobayashi}, S., {Steele}, I.~A. et al. \ 2016, GCNC, 20074
\bibitem[Guidorzi et al.(2016B)]{2016GCN.20080....1G} {Guidorzi}, C., {Kobayashi}, S., {Steele}, I.~A. et al. \ 2016, GCNC, 20080
\bibitem[Hui \& Meegan(2016)]{2016GCN..20068...1H} Hui, C.~M., \& Meegan, C.\ 2016, GRB Coordinates Network, 20068, 1 
\bibitem[Ioka et al.(2005)]{2005ApJ...631..429I} {Ioka}, K., {Kobayashi}, S. \& {Zhang}, B. \ 2005, \apj, 631, 429 
\bibitem[Fujiwara et al.(2016)]{2016Fuji} Fujiwara, T., Saito, Y., Tachibana, Y., et al. \ 2016, GCNC 
\bibitem[Freedman \& Waxman(2001)]{2001ApJ...547..922F} {Freedman}, D.~L. and {Waxman}, E., \ 2001, \apj, 547, 922
\bibitem[Kaur et al.(2016)]{2016GCN.20079....1K} {Kaur}, A., {Henson}, G. \& {Hartmann}, D.~H. \ 2016, GCNC, 20079
\bibitem[Keshet \& Waxman(2005)]{2005PhRvL..94k1102K} {Keshet}, U. \& {Waxman}, E. \ 2005, Physical Review Letters, 94, 11 
\bibitem[Kobayashi(2000)]{2000ApJ...545..807K} Kobayashi, S.\ 2000, \apj, 545, 807
\bibitem[Kobayashi(2003)]{2003ApJ...597..455K} {Kobayashi}, S. \& {Zhang}, B.\ 2003, \apj, 597, 455 
\bibitem[Kotani et al.(2005)]{2005NCimC..28..755K} Kotani, T., Kawai, N., Yanagisawa, K., et al.\ 2005, Nuovo Cimento C Geophysics Space Physics C, 28, 755 
\bibitem[Kr{\"u}hler et al.(2009)]{2009ApJ...697..758K} {Kr{\"u}hler}, T., {Greiner}, J., {McBreen}, S., et al. \ 2009, \apj, 697, 758 
\bibitem[Kumar \& Piran(2000)]{2000ApJ...532..286K} Kumar, P., \& Piran, T.\ 2000, \apj, 532, 286 
\bibitem[Lazzati et al.(2002)]{2002A&A...396L...5L} Lazzati, D., Rossi, E., Covino, S., Ghisellini, G., \& Malesani, D.\ 2002, \aap, 396, L5 
\bibitem[Li et al.(2015)]{2015ApJ...805...13L} Li, L., Wu, X.-F., Huang, Y.-F., et al.\ 2015, \apj, 805, 13 
\bibitem[Melandri et al.(2008)]{2008ApJ...686.1209M} Melandri, A., Mundell, C.~G., Kobayashi, S., et al.\ 2008, \apj, 686, 1209
\bibitem[Melandri et al.(2010)]{2010ApJ...723.1331M} {Melandri}, A., {Kobayashi}, S., {Mundell}, C.~G., et al.\ 2010, \apj, 723, 1331
\bibitem[Melandri et al.(2014)]{2014A&A...572A..55M} {Melandri}, A., {Virgili}, F.~J., {Guidorzi}, C., et al. \ 2014, \aap, 572, A55
\bibitem[Melandri et al.(2016)]{2016GCN.20071....1M} {Melandri}, A., {D'Avanzo}, P., {D'Elia}, V., et al. \ 2016, GCNC, 20071
\bibitem[M{\'e}sz{\'a}ros \& Rees(1997)]{1997ApJ...476..232M} M{\'e}sz{\'a}ros, P., \& Rees, M.~J.\ 1997, \apj, 476, 232 
\bibitem[M{\'e}sz{\'a}ros et al.(1998)]{1998ApJ...499..301M} M{\'e}sz{\'a}ros, P., Rees, M.~J., \& Wijers, R.~A.~M.~J.\ 1998, \apj, 499, 301 
\bibitem[Nardini et al.(2011)]{2011A&A...531A..39N} {Nardini}, M., {Greiner}, J., {Kr{\"u}hler}, T., et al. \ 2011, \aap, 531, A39 
\bibitem[Oates et al.(2007)]{2007MNRAS.380..270O} Oates, S.~R., de Pasquale, M., Page, M.~J., et al.\ 2007, \mnras, 380, 270 
\bibitem[Panaitescu et al.(1998)]{1998ApJ...503..314P} Panaitescu, A., M{\'e}sz{\'a}ros, P., \& Rees, M.~J.\ 1998, \apj, 503, 314 
\bibitem[Perley et al.(2008)]{2008ApJ...688..470P} {Perley}, D.~A., {Li}, W., {Chornock}, R., et al. \ 2008, \apj, 688, 470 
\bibitem[Racusin et al.(2008)]{2008Natur.455..183R} Racusin, J.~L., Karpov, S.~V., Sokolowski, M., et al.\ 2008, \nat, 455, 183 
\bibitem[Racusin et al.(2009)]{2009ApJ...698...43R} Racusin, J.~L., Liang, E.~W., Burrows, D.~N., et al.\ 2009, \apj, 698, 43 
\bibitem[Rees \& M{\'e}sz{\'a}ros(1998)]{1998ApJ...496L...1R} Rees, M.~J., \& M{\'e}sz{\'a}ros, P.\ 1998, \apjl, 496, L1 
\bibitem[Rees \& M{\'e}sz{\'a}ros(2000)]{2000ApJ...545L..73R} Rees, M.~J., \& M{\'e}sz{\'a}ros, P.\ 2000, \apjl, 545, L73 
\bibitem[Ricker et al.(2003)]{2003AIPC..662....3R} Ricker, G.~R., Atteia, J.-L., Crew, G.~B., et al.\ 2003, Gamma-Ray Burst and Afterglow Astronomy 2001: A Workshop Celebrating the First Year of the HETE Mission, 662, 3 
\bibitem[Sari et al.(1998)]{1998ApJ...497L..17S} {Sari}, R., {Piran}, T. \& {Narayan}, R.\ 1998, \apjl, 497, L17 
\bibitem[Sari \& Piran(1999a)]{1999ApJ...519L..17S} {Sari}, R., {Piran}, T. \& {Halpern}, J.~P.\ 1999, \apjl, 519, L17 
\bibitem[Sari \& Piran(1999b)]{1999ApJ...520..641S} Sari, R., \& Piran, T.\ 1999, \apj, 520, 641 
\bibitem[Sato et al.(2003)]{2003ApJ...599L...9S} Sato, R., Kawai, N., Suzuki, M., et al.\ 2003, \apjl, 599, L9 
\bibitem[Schaefer et al.(2003)]{2003ApJ...588..387S} Schaefer, B.~E., Gerardy, C.~L., H{\"o}flich, P., et al.\ 2003, \apj, 588, 387 
\bibitem[Shimokawabe et al.(2008)]{2008AIPC.1000..543S} Shimokawabe, T., Kawai, N., Kotani, T., et al.\ 2008, American Institute of Physics Conference Series, 1000, 543 
\bibitem[Tachibana et al.(2017)]{2017PASJ...69...63T} {Tachibana}, Y., {Yoshii}, T., {Hanayama}, et al., \ 2017, \pasj, 69, 63 
\bibitem[Th{\"o}ne et al.(2007)]{2007ApJ...671..628T} Th{\"o}ne, C.~C., Greiner, J., Savaglio, S., \& Jehin, E.\ 2007, \apj, 671, 628 
\bibitem[Torii et al.(2003)]{2003ApJ...597L.101T} Torii, K., Kato, T., Yamaoka, H., et al.\ 2003, \apjl, 597, L101 
\bibitem[Troja et al.(2016)]{2016GCN..20064...1T} Troja, E., Burrows, D.~N., D'Avanzo, P., et al.\ 2016, GRB Coordinates Network, 20064, 1 
\bibitem[Uemura et al.(2004)]{2004PASJ...56S..77U} Uemura, M., Kato, T., Ishioka, R., et al.\ 2004, \pasj, 56, S77 

\bibitem[Waxman (1997)]{1997ApJ...485L...5W} {Waxman}, E. \ 1997, \apjl, 485, L5
\bibitem[Yatsu et al.(2007)]{2007PhyE...40..434Y} Yatsu, Y., Kawai, N., Shimokawabe, T., et al.\ 2007, Physica E Low-Dimensional Systems and Nanostructures, 40, 434 
\bibitem[Yatsu et al.(2016)]{2016RMxAC..48...24Y} Yatsu, Y., Kawai, N., Fujiwara, T., et al.\ 2016, Revista Mexicana de Astronomia y Astrofisica Conference Series, 48, 24 
\bibitem[Zhang et al.(2006)]{2006ApJ...642..354Z} {Zhang}, B., {Fan}, Y.~Z. \& {Dyks}, J. \ 2006, \apj, 642, 354 

\end{thebibliography}
\end{document}